\documentclass[aps,prl,10pt,twocolumn,superscriptaddress,floatfix]{revtex4-2}
\usepackage{chngcntr}
\usepackage{epstopdf}
\usepackage{graphicx}
\usepackage{graphics}
\usepackage{dcolumn}
\usepackage{bm}
\usepackage{bbm,bbold}
\usepackage{color}
\usepackage{xcolor, soul}
\sethlcolor{green}
\usepackage{amsfonts}
\usepackage{amsmath}
\usepackage{mdframed}
\usepackage[normalem]{ulem}
\usepackage{mathrsfs}
\usepackage[percent]{overpic}

\usepackage{subfigure}
\usepackage{braket}
\usepackage[colorlinks=true,citecolor=blue]{hyperref}
\hypersetup{breaklinks=true,colorlinks=true,citecolor=blue,linkcolor=blue,urlcolor=blue}

\begin{document}
	\title{ Dual-cavity controllable quantum battery}
	\author{Dayang Zhang}
	\affiliation{Key Laboratory of Optical Field Manipulation of Zhejiang
Province, Department of Physics, Zhejiang Sci-Tech University, Hangzhou
310018, China}
	\author{Shuangquan Ma}
	\affiliation{Key Laboratory of Optical Field Manipulation of Zhejiang
Province, Department of Physics, Zhejiang Sci-Tech University, Hangzhou
310018, China}
	\author{Youbin Yu}
	\email{ybyu@zstu.edu.cn}
	\affiliation{Key Laboratory of Optical Field Manipulation of Zhejiang
Province, Department of Physics, Zhejiang Sci-Tech University, Hangzhou
310018, China}
	\author{Guangri Jin}
	\affiliation{Key Laboratory of Optical Field Manipulation of Zhejiang
Province, Department of Physics, Zhejiang Sci-Tech University, Hangzhou
310018, China}
	\author{Aixi Chen}
	\affiliation{Key Laboratory of Optical Field Manipulation of Zhejiang
Province, Department of Physics, Zhejiang Sci-Tech University, Hangzhou
310018, China}
\begin{abstract}
	With the rapid development of quantum science and technology, quantum batteries have also emerged. However, there are still many unresolved issues in the field of quantum batteries. For example, how to improve battery space utilization, maximize battery energy storage, and how to increase and control the charging power of quantum batteries. A major challenge is how to achieve better charging power without reducing the energy storage of the quantum batteries. Here, we propose a controllable dual-cavity quantum battery which can increase the charging power without diminishing capacity of the quantum batteries by manipulating the number of atoms . This control method can effectively adjust the charging power of quantum batteries from $N^{2}$ times to $N^{2.5}$ times, and even to $N^{3}$ times. By adjusting the number of atoms, quantum batteries can achieve theoretical ``fast charging'' and ``slow charging''.
\end{abstract}
	\maketitle
	Quantum informatics and quantum thermodynamics have gradually become the focus of attention with the development of quantum science. As an important part of quantum informatics, optical quantum information has been widely studied. Such as the preparation of squeezed state light field \cite{PhysRevX.10.021049,steinbrecher2019quantum,liu2001observation} and quantum teleportation \cite{hu2023progress,bouwmeester1997experimental,ren2017ground,sherson2006quantum,furusawa1998unconditional}, which are closely related to quantum entanglement \cite{RevModPhys.81.865,erhard2020advances} and steering \cite{RevModPhys.92.015001,PhysRevLett.114.060403,Liu_2020}. Quantum thermodynamics and quantum information are inseparable. Traditional thermodynamics cannot describe quantum scale devices and requires a new understanding of concepts such as work, heat, and entropy. This has led to new understandings of these quantities in the field of quantum thermodynamics, including research on quantum machines such as heat engines and refrigerators \cite{niedenzu2018quantum,millen2016perspective,goold2016role,martinez2017colloidal,tu2021abstract}. In the process of studying quantum information and quantum thermodynamics, people discovered the potential advantages of quantum effects in some applications, such as in quantum sensing \cite{degen2017quantum},
cryptography \cite{pirandola2020advances}, and computation \cite{divincenzo2000physical}. One scenario which features
both of these aspects of quantum science is that of the
possible quantum enhancement in thermodynamic tasks, such
as the charging of batteries \cite{alicki2013entanglement,campaioli2017enhancing,le2018spin,ferraro2018high,barra2019dissipative,pirmoradian2019aging,zhang2019powerful,andolina2019extractable,carrega2020dissipative,jiang2022quantum,PhysRevA.102.060201,PhysRevLett.132.210402,10.1116/5.0184903,PhysRevLett.132.090401,PhysRevA.109.042424,PhysRevA.109.032201,PhysRevA.109.052206,PhysRevResearch.6.023136,rosa2020ultra,PhysRevLett.128.140501,Shaghaghi_2022,PhysRevLett.125.236402,PhysRevLett.131.060402}.

The concept of quantum batteries (QBs) was first proposed by Alicki and Fannes, who showed that quantum batteries can utilize quantum entanglement to improve extractable work compared to traditional batteries \cite{alicki2013entanglement}. QBs are different from traditional batteries. It is considered a device where two-level atoms act as battery cells and other particles or optical cavities (fields) act as chargers, utilizing their interactions to achieve charging effects. Recently, Ferraro \textit{et al.} found that collective charging in QBs can increase charging power, meaning more cells requiring less charging time \cite{ferraro2018high}. Crescente \textit{et al.} showed that charging a QB with two-photons can increase the charging speed of the battery by $N^2$ times \cite{PhysRevB.102.245407}. Carrasco \textit{et al.} found that the advantages of collective charging of QBs also exists in dissipative environments \cite{carrasco2110collective}. Gymm \textit{et al.} explored the power limit of quantum batteries and showed that the currently known methods can make the maximum power reach $N^2$ times \cite{PhysRevLett.128.140501}. In these studies, the reason for the increase in QBs charging power may be quantum entanglement or quantum coherence in the system. Kamin \textit{et al.} proved that quantum coherence is an important reason for improving the efficiency of QB and other devices \cite{PhysRevE.102.052109}. Shi \textit{et al.} studied the relationship between quantum resources and extractable work in QBs, and the results showed that quantum coherence or battery charger entanglement are necessary resources for generating non-zero extractable work during the charging process \cite{PhysRevLett.129.130602}.

On the other hand, the interactions between atoms also have a significant impact on QB. Zhang \textit{et al.} used the harmonic charging field to charge the two-level atoms, showing that the repulsive force between the atoms has a positive effect on the charging power and energy storage, while the attraction has a negative effect \cite{zhang2019powerful}. Jiang \textit {et al.} confirmed this phenomenon and showed that mixed fields can improve the energy storage and the power of QBs, which is mainly due to the influence of atomic interactions \cite{jiang2022quantum}.

These studies indicate that QB has many available resources, but these resources cannot solve all problems. For example, in many rechargeable batteries, the usage rate of QB cannot reach $1 $. QB usually requires sacrificing charging power to achieve maximum storage energy, or sacrificing storage energy to increase charging power. Can one balance both, that is, achieve a larger charging power, the QB also has a larger capacity?

\begin{figure}[tb]
	\centering
	\includegraphics[width=1\linewidth]{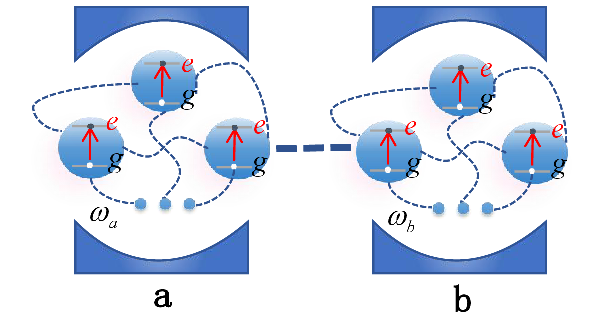}
	\caption{Two two-level atomic chains with interatomic interactions are coupled into two optical cavities, respectively. They have collective interactions between atoms.}
	\label{fig1}
\end{figure}

In this letter, to solve this problem, we propose a dual-cavity QB with controllable characteristics. It can affect the charging power and the energy storage of QB by controlling the number of atoms in the two cavities. Interestingly, we find that larger charging power and storage energy of the QBs can be achieved simultaneously by using this control method. In addition, this control method can adjust the charging power of QBs, breaking through the upper limit of ordinary methods, and the maximum charging power can reach $N^{2.5}$ times, or even higher $N^{3}$ times. It can also achieve theoretical ``fast charging'' and ``slow charging'' for the QBs.

The system consists of two optical cavities and two two-level atomic chains inside the cavity, as shown in Figure~\ref{fig1}. Because the two cavities are very close, the interaction between the two atomic chains cannot be ignored (collective interaction). In addition, we also considered the interactions between atoms in the two-level atomic chain. The Hamiltonian of the system can be written as
\begin{align}
	{H}_{Z}(t) &= {H_Q} + {H_E} + \Theta(t)H_I,\label{EQ1}
\end{align}
where ${H_Q}$ is the Hamiltonian of the two-level atoms as the capacity of the QB. ${H_E}$ denotes the Hamiltonian of a photon, and $\Theta(t)H_I$ denotes the Hamiltonian of the collective interaction atoms in the two optical cavities, as well as the interaction between the photons and the two-level atoms \cite{zhang2019powerful}. $\Theta(t)$ can be used as a switch for QB. When $\Theta(t) \neq~0$, the battery is charged by the interaction force, and when $\Theta(t) = 0$, the battery itself is discharging. The three parts of the Hamiltonian can be written as
\begin{align}
	H_Q =&\frac{\omega_{ q } }{2}\sum_{i=1}^{{N}_{a}}\sigma _{i}^{z}+\frac{g}{2N_{a}}%
	\sum_{i}^{N_{a}}(\sigma _{i}^{x}\sigma _{j}^{x}+\sigma _{i}^{y}\sigma
	_{j}^{y})  \notag \\
	&+\frac{\omega_{ q } }{2}\sum_{m=1}^{{N}_{b}}\sigma _{m}^{z}+\frac{g}{2N_{b}}%
	\sum_{m}^{N_{b}}(\sigma _{m}^{x}\sigma _{k}^{x}+\sigma _{m}^{y}\sigma
	_{k}^{y})  \notag \\
	=&\omega_{ q }S_a^z+\frac{g}{N_a}(S_a^{2}-S_{az}^{2}-\frac{N_a}{2})\notag\\
	&+\omega_{ q }S_b^z+\frac{g}{N_b}(S_b^{2}-S_{bz}^{2}-\frac{N_b}{2}),
 \label{2}
\end{align}
\begin{align}
	H_E=\omega_{ a } {a}_1^{\dagger} {a_1}+\omega_{ b } {a}_2^{\dagger} {a_2},
\end{align}
\begin{align}
	H_I=&g_1 S_a^x({a}_1^{\dagger}+ {a_1})+g_2 S_b^x({a}_2^{\dagger}  +{a_2})\notag\\&+A(S_a^{+}S_b^{-}+S_b^{+}S_a^{-}),
\end{align}
where $\sigma_{i}$ ($\sigma_{m}$) is the Pauli operator for the $i$th ($m$th) atom of cavity \textbf{a} (\textbf{b}). $\omega_q$ is the frequency of spins. $g$ is the coupling constant between atoms. $S_a=\frac{1 }{2}\sum_{i=1_a}^{{N}_{a}}\sigma _{i}$ and $S_b=\frac{1 }{2}\sum_{i=1_a}^{{N}_{b}}\sigma_{m}$. Operator ${a_1}$ (${a_2}$) annihilates
a photon with the cavity field frequency $\omega_a$ ($\omega_b$) and the strength of the spin-cavity coupling is given by a dimensionless parameter $g_1$ ($g_2$). $A$ is the collective action constant of the two atomic chains. Here we focus on the resonance
regime, i.e., $\omega_q =\omega_a =\omega_b=1$, to ensure the maximum energy transfer since the weak coupling between the two-level atoms and the optical cavity in the non-resonance region has very little effect on QB.

We describe the charging process as a two-level system transitioning from an initial ground state to an excited state under the influence of the light field and its own interactions. During this process, the energy of photons is absorbed. The initial state of the system is given as
\begin{align}
	|\psi^{(N)}(0)\rangle\!=\!(|N_a\rangle \underbrace{|{ g_1}, {g_1}, \dots, { g_1}\rangle}_\text{$N_a$})\!\otimes\!(|N_b\rangle \underbrace{|{ g_2}, {g_2}, \dots, { g_2}\rangle}_\text{$N_b$}),
	\label{5}
\end{align}
where $N=N_a+N_b$ denotes the total number of two-level atoms.
The system satisfies the time evolution equation, and the wave function at time $t$ can be written as
 \begin{align}
	|\psi^{(N)}(t)\rangle\!=\!e^{-i\int_{0}^{t} {H}_{Z}(t)dt}|\psi^{(N)}(0)\rangle.
\end{align}

Here, the two-level atoms are used as the energy storage system for QBs and the storage energy is given by the following equation \cite{ferraro2018high,jiang2022quantum}
\begin{align}
	E(t)\!=\!\langle\psi^{(N)}(t)|H_Q|\psi^{(N)}(t)\rangle\!-\!\langle\psi^{(N)}(0)|H_Q|\psi^{(N)}(0)\rangle.
	\label{7}
\end{align}

The charging power of the QB is $P=E(t)/t$. It is important to choose an appropriate time to stop evolution in order to maximize the extractable power. Therefore, the maximum stored energy $E_{max}$ (at time $t_{E}$) obtained from a given QB can be quantified as \cite{dou2022cavity}
\begin{align}\label{Emax}
	E_{max}\equiv \mathop{max}\limits_{t}[E(t)]=E[(t_{E})],
\end{align}
and the accordingly maximum charging power $P_{max}$ (at time $t_{P}$) reads
\begin{align}\label{Pmax}
	P_{max}\equiv \mathop{max}\limits_{t}[P(t)]= P[(t_{P})].
\end{align}

The system can be represented by a set of bases $|n,j,m\rangle$ to reduce the dimensionality of space, where $n$ represents the number of photons. $j$ is the angular momentum quantum number and $m$ is the magnetic quantum number. Then, the initial state in the Eq.(\ref{5}) reads $|\psi^{(N)}(0)\rangle=|N_a,\frac{N_a}{2},-\frac{N_a}{2}\rangle\otimes|N_b,\frac{N_b}{2},-\frac{N_b}{2}\rangle$. The corresponding calculation relationships for the bases are \cite{dou2022cavity}
\begin{align}
	&{a}^{\dagger}|n,j,m\rangle=\sqrt{n+1}|n+1,j,m\rangle,\notag\\
	&{a}|n,j,m\rangle=\sqrt{n}|n-1,j,m\rangle,\\
	&{J}_{\pm}|n,j,m\rangle = \sqrt{j(j+1)-m(m\pm1)}|n,j,m\pm1\rangle.\notag
\end{align}

In this QB system, the theoretical maximum limit of storage energy is as follows: the energy of all photons is absorbed by two-level atoms, which causes the two-level atoms to be in a completely inverted state at the final moment (at time $\tau$), i.e. $|\psi^{(N)}(\tau)\rangle=|0,\frac{N_a}{2},\frac{N_a}{2}\rangle\otimes|0,\frac{N_b}{2},\frac{N_b}{2}\rangle$. One can calculate the maximum limit of storage energy in the QB by
\begin{align}
	E(\tau)=&\langle\psi^{(N)}(\tau)|H_Q|\psi^{(N)}(\tau)\rangle- \langle\psi^{(N)}(0)|H_Q|\psi^{(N)}(0)\rangle\notag\\=&N\omega_{ q }.
 \label{11}
\end{align}

 From Eq.(\ref{11}), it can be seen that the maximum capacity of QB is directly proportional to total number of atoms $N$. The maximum capacity of a single QB is $E(\tau)/N=\omega_{ q }$, which is the same as the capacity of a classic battery.

 We can further analyze this phenomenon by dividing Eq.(\ref{2}) into two parts. One part is the two-level atom itself, and the other part is the interaction between two-level atoms. We find that the interaction part does not contribute to the maximum energy storage limit (capacity) of QB, and the value obtained by acting on the completely inverted quantum state $|\psi^{(N)}(\tau)\rangle$ is 0. This is consistent with the actual situation, as two-level atoms themselves are energy storage units. Their interaction can only have some positive or negative effects during the charging process, and has no effect on the maximum capacity of the QBs.
	\begin{figure}[tb]
	\centering
	\includegraphics[width=0.85\linewidth]{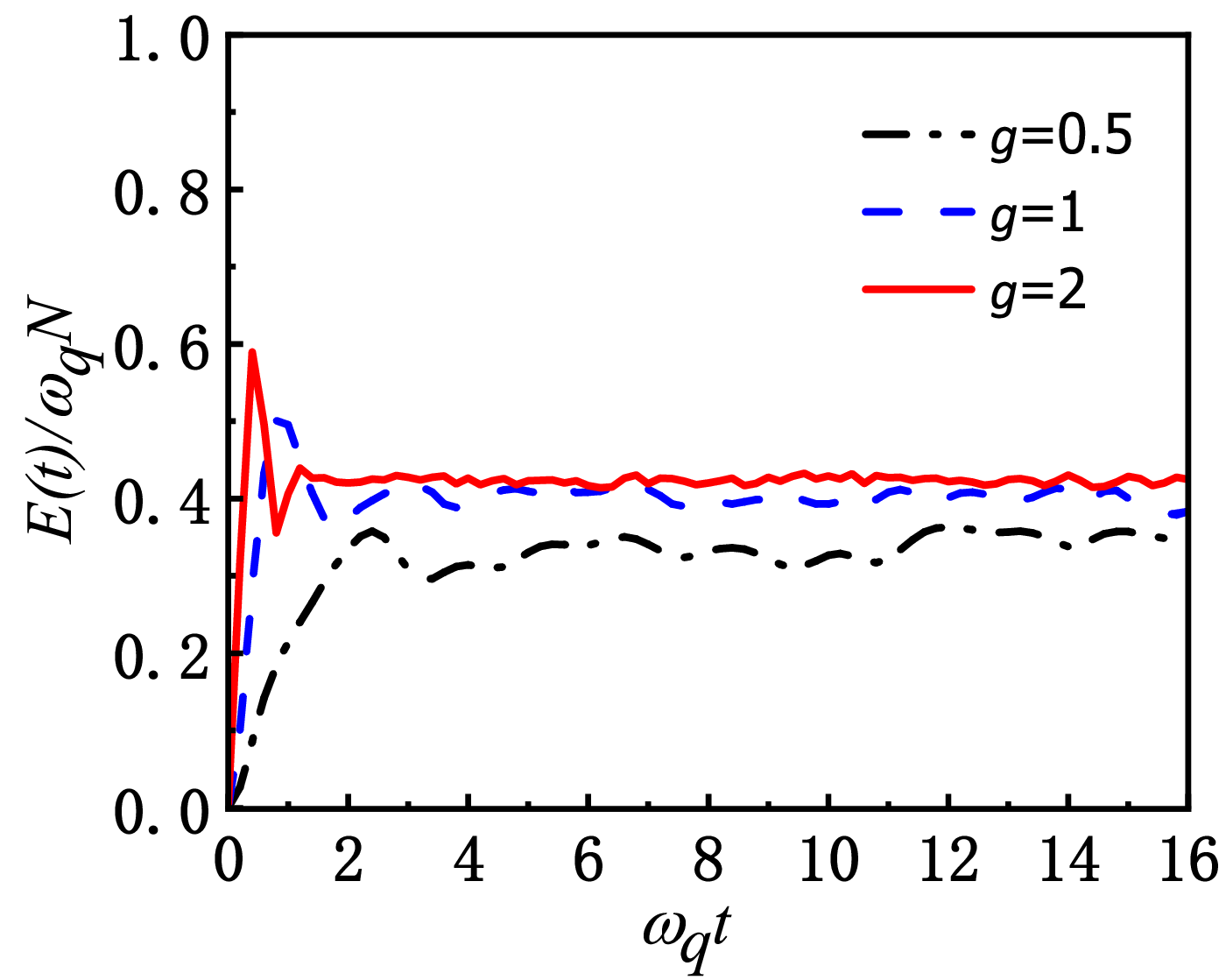}
	\caption{$E(t)/\omega_{ q }N$ versus the charging time $t$ with $g_1=g_2=0.5$ for $g=0.5, 1, $ and $2$, respectively. All shown data have been computed by setting $N = 10$ and $N_a=N_b=5$. }
	\label{fig2}
\end{figure}
\begin{figure}[tb]
	\centering
	\includegraphics[width=0.85\linewidth]{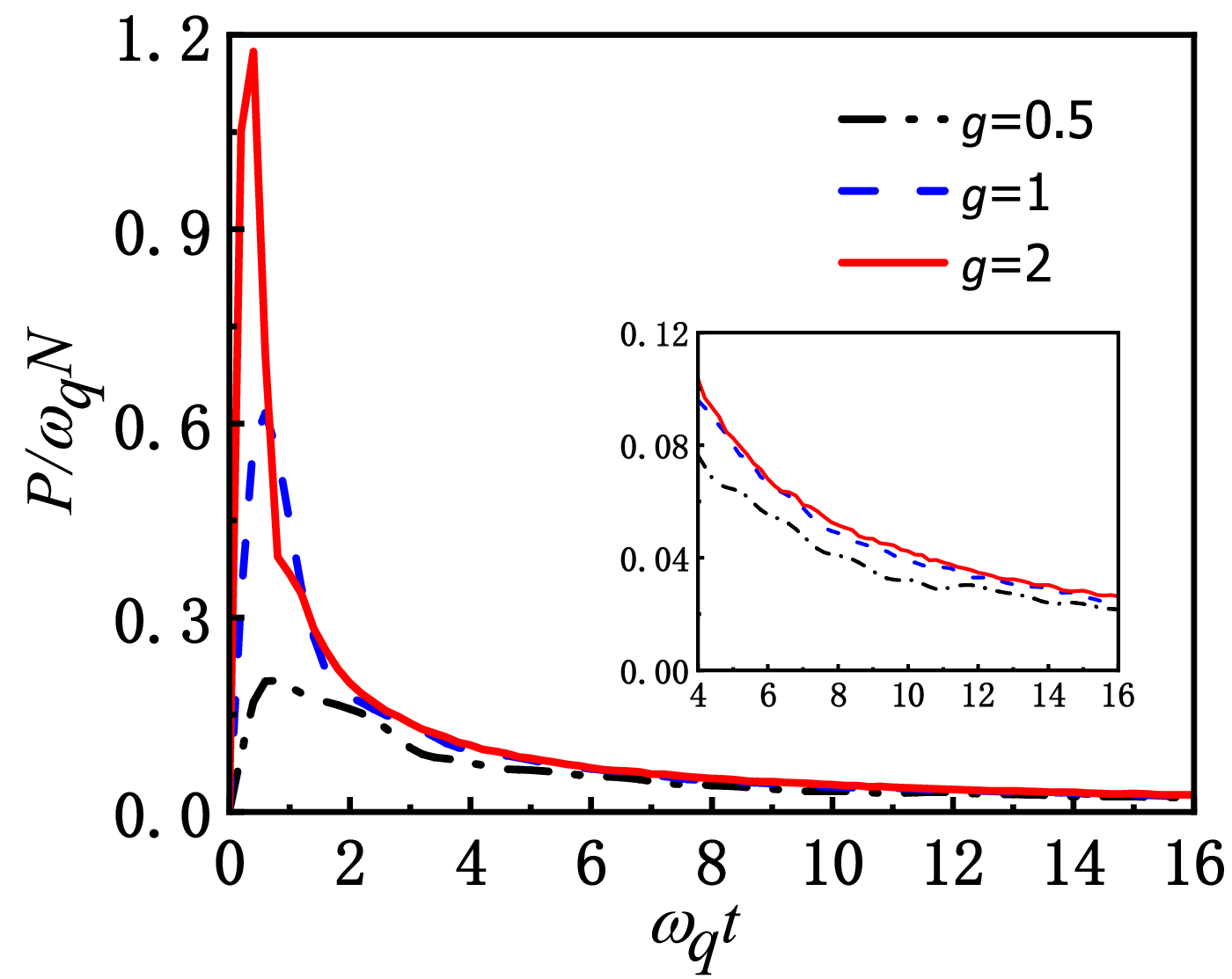}
	\caption{$P/\omega_{ q }N$ versus the charging time $t$ with $g_1=g_2=0.5$ for $g=0.5, 1$, and $2$, respectively. All shown data have been computed by setting $N = 10$ and $N_a=N_b=5$. The illustration shows an enlarged portion of time between 4 and 16. }
	\label{fig3}
\end{figure}
\begin{figure}[th]
	\centering
	\includegraphics[width=0.85\linewidth]{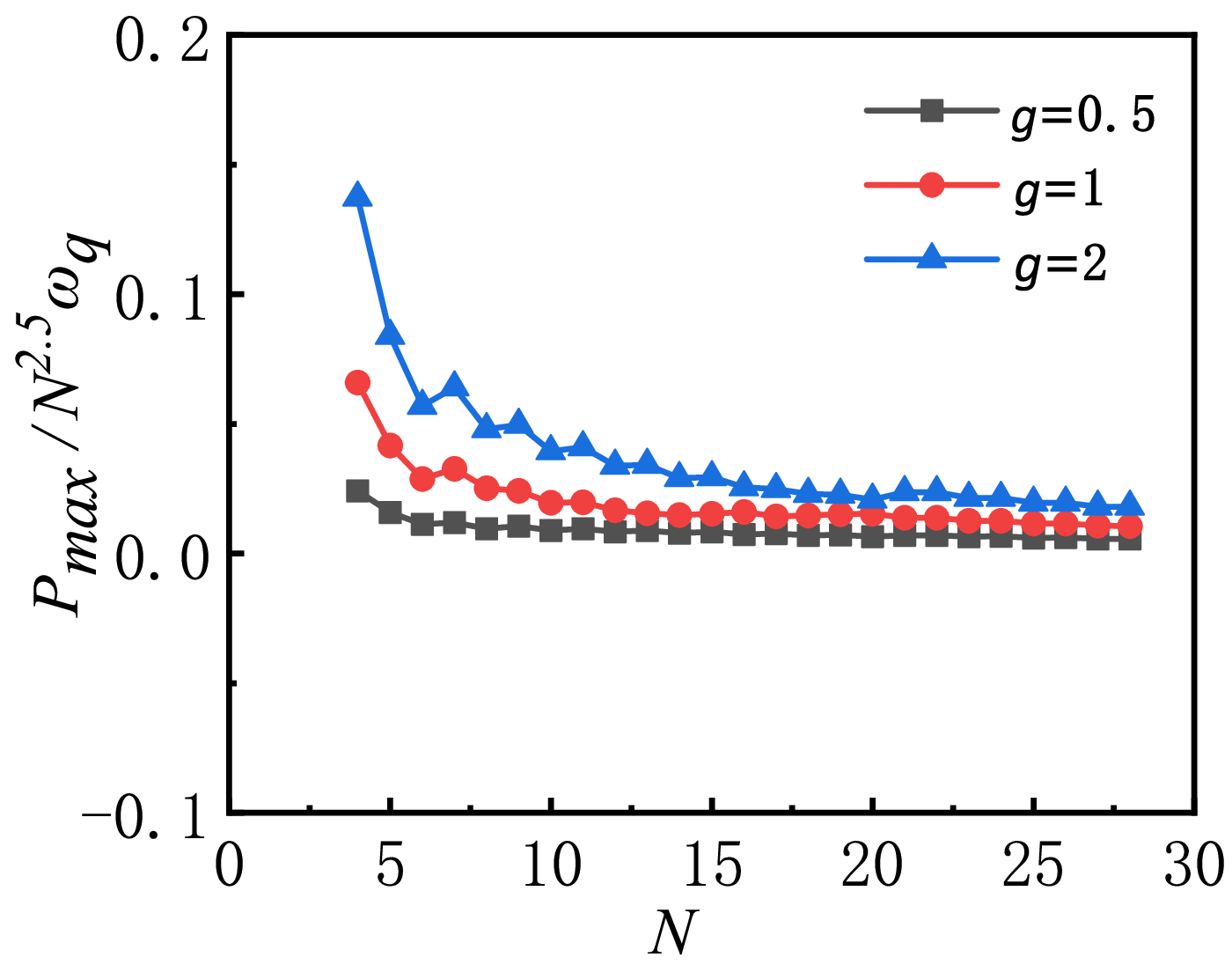}
	\caption{QB charging maximum power $P_{max}/N^{2.5}\omega_{ q }$ versus the number of atoms $N$ with $g_1=g_2=0.5$ for $g=0.5, 1, $ and $2$, respectively.}
	\label{fig4}
\end{figure}

In addition, the number of photons in the dual-cavity Hamiltonian is not conserved. There is no upper limit to the number of photons, so it can take any large integer value. In the actual calculation process, the number of photons in a finite dimensional space needs to take a maximum value of $N_{ph}$ with $N_{ph}>N$. This allows us to choose a larger $N$ value to calculate the energy stored in the QBs. In this study, we select the maximum number of photons as $N_{ph} = 30$.
	
The stored energy of a battery cannot exceed the maximum limit 1. One can see from Figure~\ref{fig2}, the interaction force between the two-level atoms can affect the energy storage of QB in the charging processes. The energy stored by the battery increases with the increase of repulsive force. But when the QB is fully charged, the effect of the interaction force is not significant. This is consistent with the previous analysis, as the interaction force has no effect on the maximum capacity of the QBs.

	\begin{figure}[tbh]
	\centering
	\includegraphics[width=0.85\linewidth]{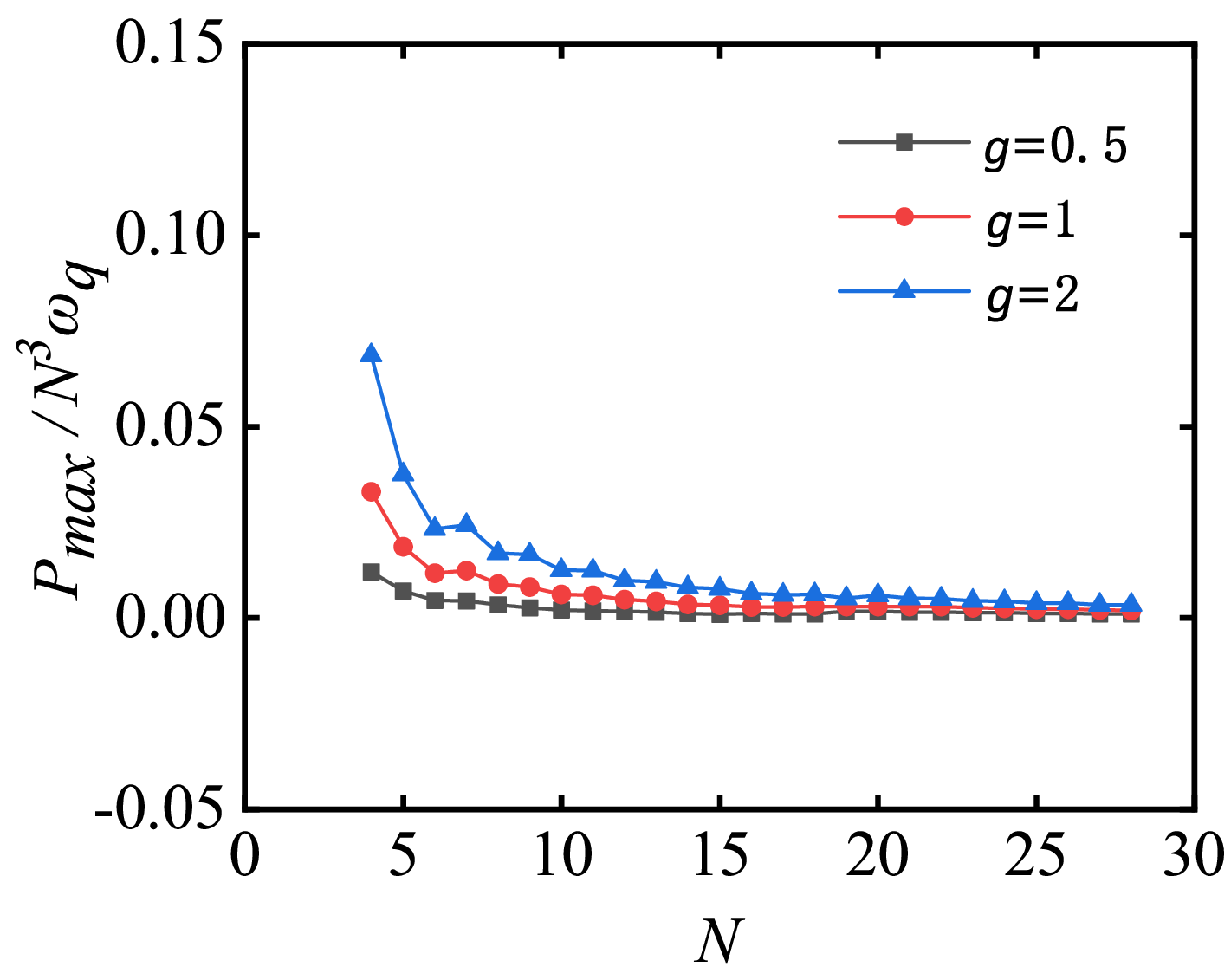}
	\caption{QB charging maximum power $P_{max}/N^{3}\omega_{ q }$ versus the number of atoms $N$ with $g_1=g_2=0.5$ for $g=0.5, 1, $ and $2$, respectively.}
	\label{fig5}
\end{figure}

Figure~\ref{fig3} is plotted to get more information about the charging power of the QBs. It is can be seen that the maximum power of QB will quickly appear. As the battery storage energy fluctuates, the power will also fluctuate. When the repulsive force between atoms increases, the charging power will increase. This indicates that the interaction force between atoms is an important reason for the increase in charging power.

The advantage of QBs collective charging is that it can amplify the charging power of the battery to $N^{1.5}$ times \cite{ferraro2018high}, or even higher to $N^{2}$ times \cite{PhysRevLett.128.140501,PhysRevB.102.245407}. Figure~\ref{fig4} shows that the curve approaches a horizontal straight line which indicates that the charging power of the QB is proportional to the square of the number of atoms $N^{2.5}$. That is to say, the charging power of our proposed QB model is $N^{2.5}$ times that of a regular battery.

We then plot Figure~\ref{fig5} by adjusting the number of the atoms in the cavities \textbf{a} and \textbf{b}. That is, to ensure that the total number of atoms is the same as in Figure~\ref{fig4}, adjust the distribution of atoms in the two cavities. Interestingly, the charging power of the QB increased by 0.5 after adjusting, reaching $N^{3}$ times. This indicates that when the number of the QBs is the same, the motion of atoms at the two cavities will form different atomic chains, leading to a change in the charging power of the QB. Therefore, we can manipulate the number of atoms to achieve control over the maximum charging power of the QBs. In addition, theoretically higher charging power can be achieved by adjusting the number of atoms in different cavities. In a sense, this can achieve two ways of charging QBs: ``fast charging'' and ``slow charging'', where the charging speed can be freely selected.

\begin{figure}[tbp]
	\centering
	\includegraphics[width=0.85\linewidth]{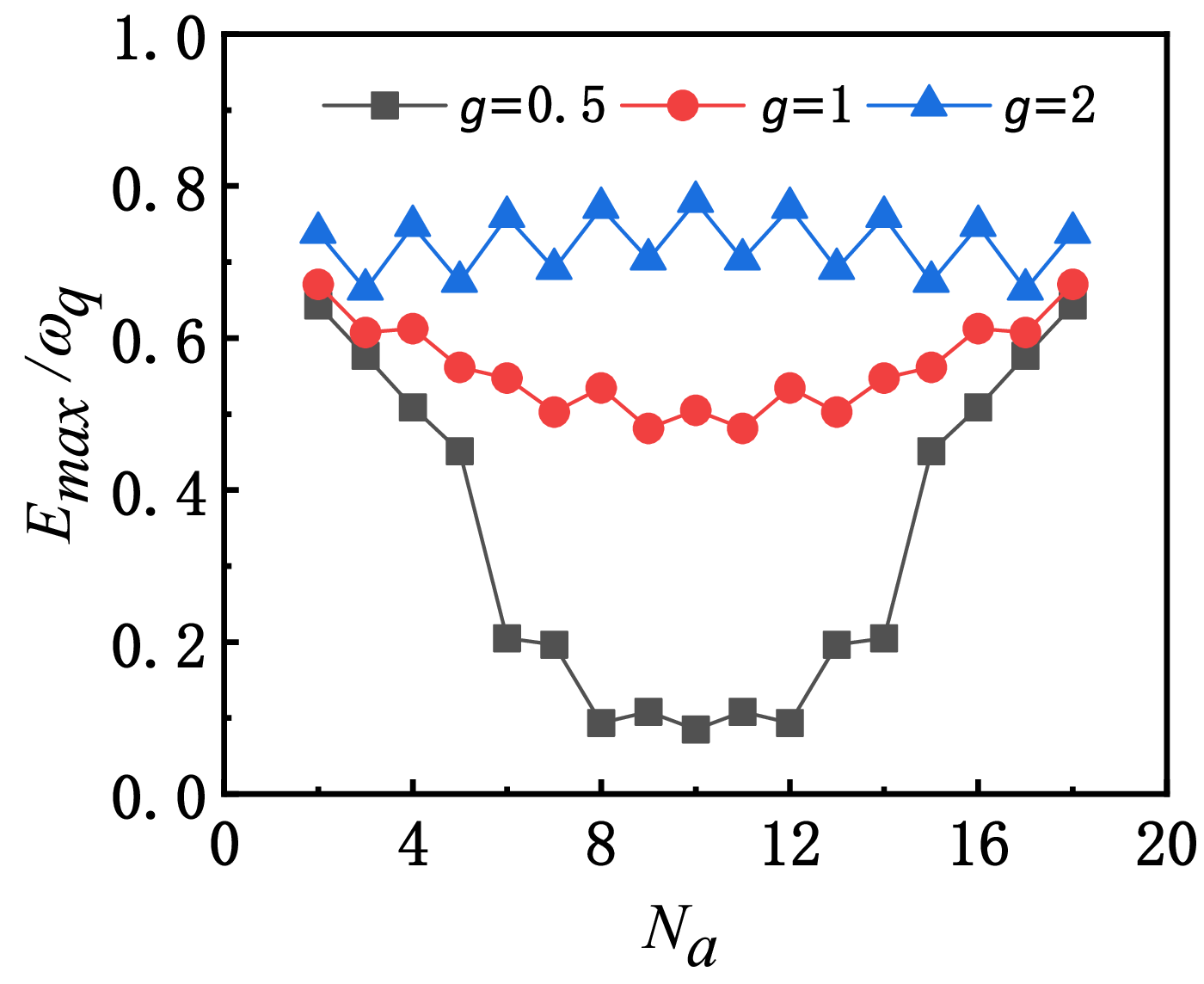}
	\caption{QB charging maximum energy storage $E_{max}/\omega_{ q }$ versus the number of atoms in cavity a $N_a$ with $g_1=g_2=0.5$ for $g=0.5, 1, $ and $2$, respectively. All shown data have been computed by setting $N=20$.}
	\label{fig6}
\end{figure}

\begin{figure}[tbp]
	\centering
	\includegraphics[width=0.85\linewidth]{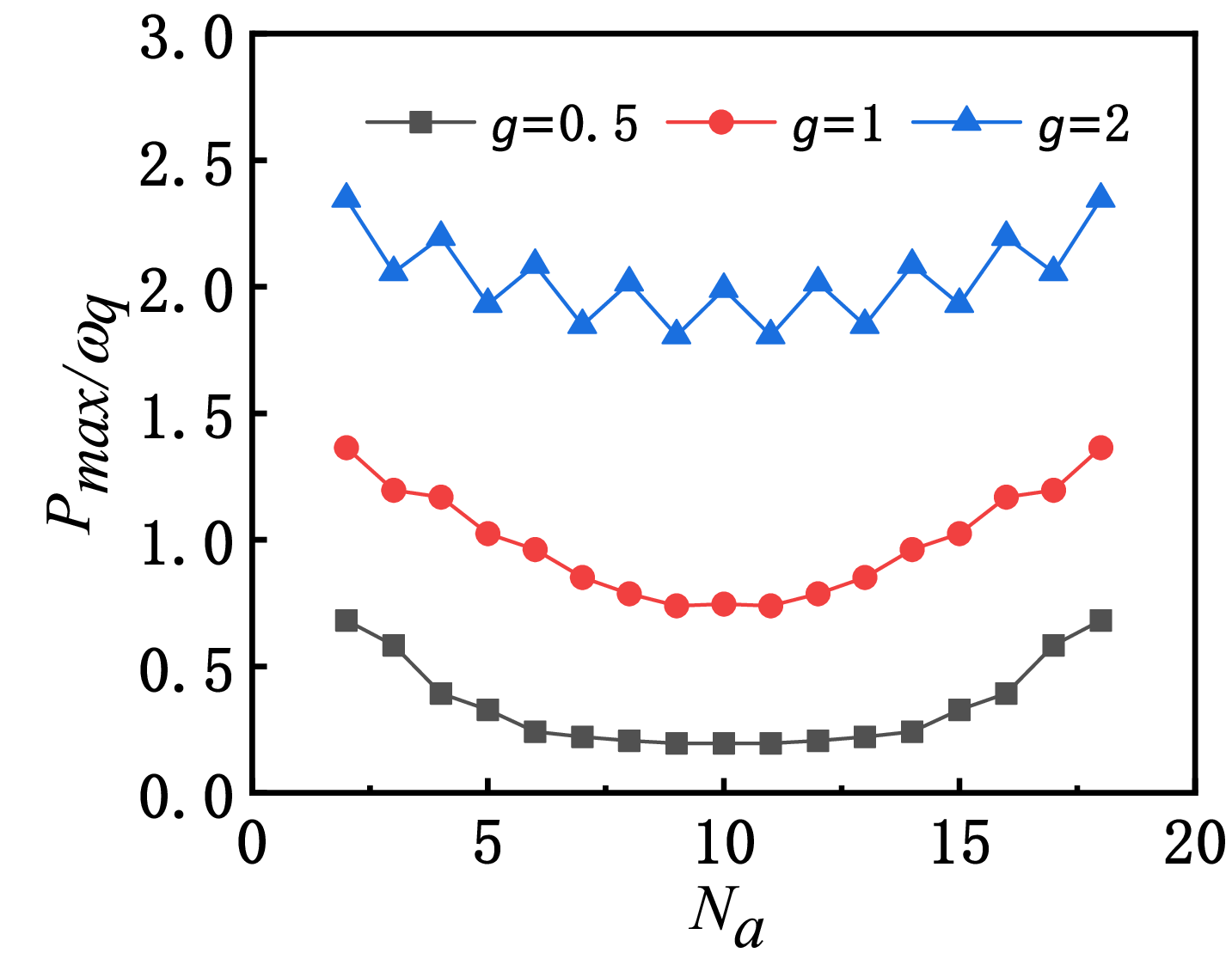}
	\caption{QB charging maximum power $P_{max}/\omega_{ q }$ versus the number of atoms in cavity a $N_a$ with $g_1=g_2=0.5$ for $g=0.5, 1, $ and $2$, respectively. All shown data have been computed by setting $N=20$. }
	\label{fig7}
\end{figure}

In order to highlight the advantages of dual-cavity controllable QBs, we further plot Figures~\ref{fig6} and~\ref{fig7}. From Figure~\ref{fig6}, one can see that the energy storage of the QB can be effectively changed by adjusting the number of atoms in cavity \textbf{a} and cavity \textbf{b} when the number of rechargeable atoms is kept the same for the interaction between atoms is relatively small. When the interaction between atoms is strong, controlling the number of atoms in two cavities has little effect on energy storage.

Figure~\ref{fig6} shows that when $g=0.5, 1$ the storage energy of the QBs first decreases and then increases with the increase of the atoms number $N_a$ in the cavity \textbf{a}. The curve is symmetrical, and when the number of atoms in two cavities is equal, the energy storage of the QBs is minimized. This may be because when the number of atoms in two cavities is the same, the system is in a relatively balanced state, with the minimum energy and the minimum amount of energy that can be stored. When the number of atoms in two cavities differs the most (We set the total number of atoms $N=20$, and
there should be at least two atoms in a cavity to achieve
collective charging.), the QB has the maximum energy storage, and the influence of atomic interactions is the weakest at this time.

Figure~\ref{fig7} shows the variation of the maximum charging power with the number of atoms in cavity \textbf{a}. From the figure, it can be seen that as the number of atoms in cavity a increases, the maximum charging power of the QB first decreases and then increases. Similar to Figure~\ref{fig6}, the charging power of the QB reaches its maximum value when the number of atoms in the two cavities differs the most.
That is to say, by adjusting the number of atoms in both cavities, QB's charging power and energy storage can have maximum values simultaneously. This breaks through the difficult problem of not being able to balance the two under normal circumstances.

In this Letter, we propose a novel dual-cavity controllable QB, which can control the maximum storage energy and maximum power of the QB by manipulating the number of atoms in each cavity. It is very meaningful that a controllable dual-cavity QB can increase its charging power by more than $N^{3}$ times through adjusting different numbers of atoms in the two cavities. Compared with sacrificing QB capacity to increase the charging power of the QB in the past, our proposed QB can suppress the decrease of QB capacity by manipulation while increasing charging power, thus obtaining larger energy storage capacity and charging power simultaneously. By adjusting the number of atoms appropriately, one can obtain the required charging power at different charging stages which provides charging modes of ``fast charging'' and ``slow charging''. This dual-cavity controllable QB offered a new means for manipulating the capacity and charging power of QBs. Especially when the interatomic interactions are relatively weak, this regulatory effect is more pronounced.

The manipulation of atoms is feasible in physical experiments. For example, one can use atomic force microscopy \cite{custance2009atomic} to achieve vertical atomic exchange \cite{sugimoto2008complex} and horizontal manipulation of copper~(111) surface \cite{stroscio2004controlling}. Complex patterning, generation, manipulation, and characterization of molecules \cite{pavlivcek2017generation} can also be achieved by adjusting the polarity of charge carriers in N-heterocyclic carbene single molecule bonds \cite{Wang_2020}. Dual-cavity controllable QBs can be controlled by manipulating atoms through this technology.
\begin{acknowledgments}
	This work is supported by National Natural Science Foundations of China (Nos. 61975184, 12175199, 12075209), Science Foundation of Zhejiang Sci-Tech University (Nos. 19062151-Y, 18062145-Y).
\end{acknowledgments}

\end{document}